\documentstyle[aps,prl,epsf,floats,twocolumn]{revtex}

\newcommand{\rt}{\rightarrow}
\def\Journal#1#2#3#4{{#1} {\bf #2}, #3 (#4)}

\def\NIM{\em Nucl. Instrum. Methods}

\def\NPB{{\em Nucl. Phys.} B}
\def\PLB{{\em Phys. Lett.}  B}
\def\PRL{\em Phys. Rev. Lett.}
\def\PRD{{\em Phys. Rev.} D}

\def\NPA{{\em Nucl. Phys.} A}

\long\def\abstract#1{
        \begin{center}
        \begin{minipage}{5.2truein}
        \footnotesize
        \parindent1em\noindent\ignorespaces#1
        \end{minipage}
        \end{center}
        \vskip1.5em}

\begin{document}
\title{\boldmath Measurements of the Cross Section for $e^+e^-\rt
\mbox{hadrons}$ at Center-of-Mass Energies from 2 to 5 GeV}

\author{
J.~Z.~Bai$^1$,      Y.~Ban$^{10}$,      J.~G.~Bian$^1$,
A.~D.~Chen$^1$,     H.~F.~Chen$^{16}$,  H.~S.~Chen$^1$,
J.~C.~Chen$^1$,     X.~D.~Chen$^1$,     Y.~B.~Chen$^1$,
B.~S.~Cheng$^1$
% ...
\\(BES Collaboration)\\} 

\address{
$^1$ Institute of High Energy Physics, Beijing 100039, People's Republic of
     China\\
$^2$ China Center of Advanced Science and Technology, Beijing 100080,
     People's Republic of China\\
$^3$ Chonbuk National University, Chonju 561-756, Korea\\
$^4$ Colorado State University, Fort Collins, Colorado 80523\\
$^5$ Henan Normal University, Xinxiang 453002, People's Republic of China\\
$^6$ Huazhong Normal University, Wuhan 430079, People's Republic of China\\
$^7$ Hunan University, Changsha 410082, People's Republic of China\\
$^8$ Korea University, Seoul 136-701, Korea\\
$^{9}$ Nankai University, Tianjin 300071, People's Republic of China\\
$^{10}$ Peking University, Beijing 100871, People's Republic of China\\
$^{11}$ Shandong University, Jinan 250100, People's Republic of China\\
$^{12}$ Shanghai Jiaotong University, Shanghai 200030, 
        People's Republic of China\\
$^{13}$ Seoul National University, Seoul 151-742, Korea\\
$^{14}$ Stanford Linear Accelerator Center, Stanford, California 94309\\
$^{15}$ University of Hawaii, Honolulu, Hawaii 96822\\
$^{16}$ University of Science and Technology of China, Hefei 230026,
        People's Republic of China\\
$^{17}$ University of Texas at Dallas, Richardson, Texas 75083-0688\\
$^{18}$ Wuhan University, Wuhan 430072, People's Republic of China\\
$^{19}$ Zhejiang University, Hangzhou 310028, People's Republic of China
}

\date{\today}

\maketitle

\twocolumn[\maketitle\abstract{
We report values of $R = \sigma(e^+e^-\rightarrow
\mbox{hadrons})/\sigma(e^+e^-\rightarrow\mu^+\mu^-)$ for 85
center-of-mass energies between 2 and 5 GeV measured 
with the upgraded Beijing Spectrometer at the Beijing 
Electron-Positron Collider. 
}]


In precision tests of the Standard Model (SM)~\cite{rlowe}, 
the quantities
$\alpha(M_Z^2)$, the QED running coupling constant evaluated at
the $Z$ pole, and $a_\mu = (g-2)/2$, the anomalous magnetic moment
of the muon, are of fundamental importance.
%
%
The dominant uncertainties in both $\alpha(M^2_{Z})$ and $a_{\mu}^{SM}$ are 
due to the effects of hadronic vacuum polarization, which cannot be 
reliably calculated in the low energy region.  Instead, with the 
application of dispersion relations,
experimentally measured $R$ values are used to determine the vacuum 
polarization, where $R$ is the lowest order cross section 
for $e^+e^-\rightarrow\gamma^*\rightarrow hadrons$
in units of the lowest-order QED cross section for
$e^+e^- \rightarrow \mu^+\mu^-$, namely
$R=\sigma(e^+e^- \rightarrow \mbox{hadrons})/\sigma(e^+e^-\rightarrow
\mu^+\mu^-)$, where 
$\sigma (e^+e^- \rightarrow \mu^+\mu^-) = \sigma^0_{\mu \mu}=
4\pi \alpha^2(0) / 3s$.

Values of $R$ in the center-of-mass (c.m.) energy ($E_{cm}$) range below 5 GeV  
were measured about 20 years ago with a precision of $15 - 20\%$
~\cite{mark1,gamma2,pluto}. 
In this paper, we report measurements 
of $R$ at 85 c.m. energies between 2 and 4.8 GeV, with an average precision
of $6.6\%$ ~\cite{rscan2}.  
The measurements were carried out with the upgraded Beijing Spectrometer
(BESII)~\cite{bes2} at the Beijing Electron-Positron Collider (BEPC). 

Experimentally, the value of $R$ is determined from the number of 
observed hadronic events, $N^{obs}_{had}$, by the relation
\begin{equation}
R=\frac{ N^{obs}_{had} - N_{bg} - \sum_{l}N_{ll} - N_{\gamma\gamma} }
{ \sigma^0_{\mu\mu} \cdot L \cdot \epsilon_{trg} \cdot \bar \epsilon_{had}
\cdot (1+\delta)},
\end{equation}
where $N_{bg}$ is the number of beam-associated background events;
$\sum_{l}N_{ll},~(l=e,\mu,\tau)$ are the numbers
of lepton-pair events from one-photon processes and $N_{\gamma\gamma}$
the number of two-photon process
events that are misidentified as hadronic events;
$L$ is the integrated luminosity; $\delta$ is
the effective initial state radiative (ISR) correction; 
$\bar\epsilon_{had}$ is the average detection efficiency 
for hadronic events; and $\epsilon_{trg}$ is the trigger efficiency. 
The triggers and the integrated luminosity measurement
 were the same as those used in
a preliminary scan that measured $R$ at 6 energy points
between 2.6 and 5 GeV \cite{besr_1}.

\begin{figure}[htb]
\epsfxsize=3.4in
\centerline{\epsfbox{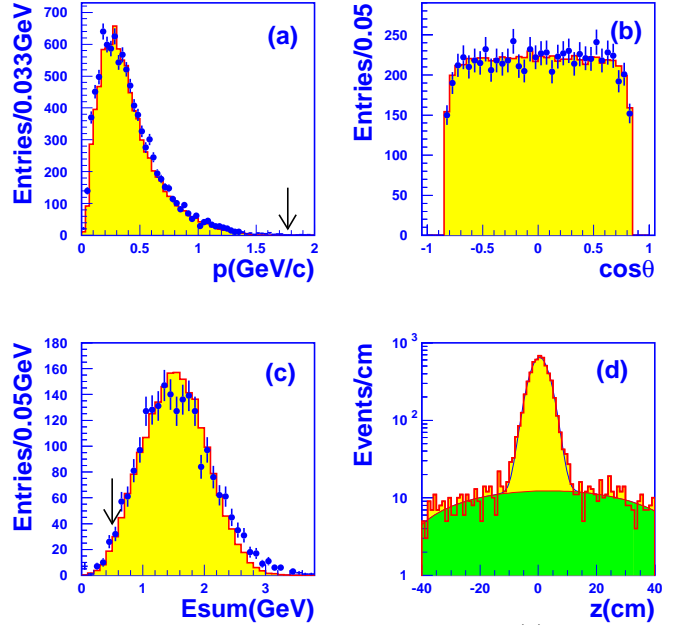}}
\caption{Distributions for $E_{cm}$=3.0 GeV of 
(a) track momentum;
(b) track $\cos \theta$; 
(c) total energy deposited in the BSC; and  
(d) event vertex position along the beam ($z$) axis. 
Histograms and dots in (a)-(c) represent Monte Carlo and real data, 
respectively; the beam associated background in (c) has been removed by
sideband subtraction.}
\label{fig:evtsel}
\end{figure}

The hadronic event selection is similar with that used in the first $R$
scan \cite{besr_1} but with improvements that include: for good 
track selection, the distance of closest approach requirement ($<18$cm) 
of a track to the interaction point along the beam axis is not imposed;
for event-level selection, the selected tracks must not all point into the 
forward ($\cos \theta > 0$) or the backward ($\cos \theta < 0$) hemisphere.
Some distributions comparing data and Monte Carlo data
are shown in Figs.~\ref{fig:evtsel} (a)-(c).
The cuts used for selecting hadronic events were varied over a 
wide range, e.g. $|\cos \theta|$ from 0.75 to 0.90, 
$E_{sum}$ from 0.24$E_{beam}$ to 0.32$E_{beam}$ ($E_{sum}$ is the total
deposited energy, $E_{beam}$ the beam energy) to estimate the 
systematic error arising from the event selection; this is the dominant 
component of the systematic error as indicated in Table~\ref{tab:error}.  

The numbers of hadronic events and beam-associated background events
are determined by fitting the distribution of event vertices along the 
beam direction with a Gaussian to describe the hadronic events and a 
polynomial of degree one to three for the beam-associated background.
This background varies from 3 to 10\% 
of the selected hadronic event candidates, depending on the energy.    
The fit using a second degree polynomial, shown in Fig.~\ref{fig:evtsel} (d),
turned out to be the best.  The difference between 
using a polynomial of degree one or three 
to that of degree two is about 1\%, which is included in the systematic 
error in the event selection.    


A special joint effort was made by the Lund 
group and the BES collaboration to develop the LUARLW generator,
which uses a formalism based on the 
Lund Model Area Law, but without the extreme-high-energy 
approximations used in JETSET's string fragmentation algorithm
~\cite{bo}. 
The final states simulated in LUARLW are exclusive, 
in contrast to JETSET, where they are inclusive. 
%
%
Above 3.77 GeV, 
the production of $D,~D^*,~D_s,$ and $D_s^*$
is included in the generator according to the Eichten 
Model~\cite{eichiten}.
A Monte Carlo event generator has been developed
to handle decays of the resonances in the radiative return processes
$e^+e^- \rightarrow \gamma J/\psi$ or $\gamma \psi(2S)$~\cite{chenjc}.

\begin{figure}[htb]
\epsfysize=2.7in
\centerline{\epsfbox{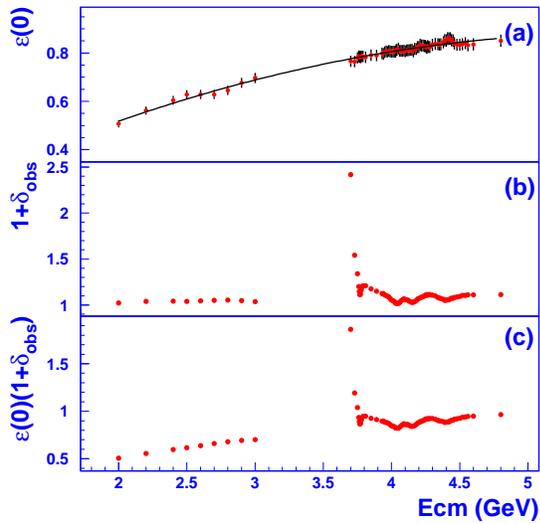}}
\caption{(a) The c.m. energy dependence of the detection efficiency
for hadronic events estimated using the LUARLW generator. The error 
bars are the total systematic errors. 
(b) The calculated radiative correction, and
(c) the product of (a) and (b).}
\label{fig:eff_isr99}
\end{figure}

The parameters in LUARLW are tuned to reproduce
14 distributions of kinematic variables 
over the entire energy region covered by the scan~\cite{huhm}.
We find that one set of parameter values is required for the c.m. energy
region below open charm threshold, and that a second set
is required for higher energies.
In an alternative approach, the parameter values 
were tuned point-by-point throughout the
entire energy range.  
The detection efficiencies 
determined using individually tuned parameters are consistent with 
those determined with globally tuned parameters to within 2\%.   
This difference is included in the systematic errors. 
The detection efficiencies were also determined using JETSET74 for 
the energies above 3 GeV. 
The difference between the JETSET74 and LUARLW results is about 1\%,
and is also taken into account in estimating the systematic 
uncertainty.  Figure~\ref{fig:eff_isr99}~(a) shows the variation 
of the detection efficiency as a function of c.m. energy. 

We changed the 
fractions of $D,~D^*,~D_s,$ and $D_s^*$ production by 50\% and find that 
the detection efficiency varies less than 1\%.  We also varied 
the fraction of the continuum under the broad resonances by 20\%, and find
the change of the detection efficiency is about 1\%. These 
variations are included in the systematic errors. 

Different schemes for the initial state radiative corrections
were compared\cite{berends,fmartin,fadin,crystalball},
as reported in ref. \cite{besr_1}. Below charm threshold, 
the four different
schemes agree with each other to within 1\%, while
above charm threshold, where resonances are
important, the agreement is within 1 to 3\%.  
The radiative correction used in this analysis is based on 
ref.~\cite{crystalball},
and the differences with the other schemes are included in the
systematic error \cite{comment}.
In practice, the radiative effects in the detection efficiency
were moved into radiative correction factor by making the
replacement $\bar\epsilon_{had}(1+\delta) \to \epsilon(0) 
(1+\delta_{obs})$, where 
$\epsilon(k)$ is the efficiency
for events with a radiative photon of energy $k$, and 
$\delta_{obs}$ contains a modification of the bremsstrahlung
term to reflect the $k$-dependence of the hadronic acceptance.

To calculate $\delta_{obs}$, 
a cutoff in 
$s'$, the effective c.m. energy after ISR to produce hadrons, 
has to be made. In our calculation, the mimimum value of $s'$
should be the threshold for producing two pions, corresponding to
$k_{max}=1-s'/s=(0.9805-0.9969)$ in the 2-5 GeV range.
Our criteria to select hadronic events is such that $\epsilon$ 
approaches zero when $k$ is close to 0.90, which makes us 
insensitive to events with high ISR photon energy. 

In calculating the radiative correction for the narrow resonances
$J/\psi$ and $\psi(2S)$, the theoretical cross 
section is convoluted 
with the energy distribution of the 
colliding beams, which is treated as a Gaussian with 
a relative beam energy spread of $1.32 \times 10^{-4}~E_{cm}$ ($E_{cm}$ in GeV).
For the broad resonances 
at 3770, 4040, 4160, and 4416 MeV, the interferences and the 
energy-dependence of total widths were taken into consideration.
Initially
the resonance parameters from PDG2000~\cite{PDG2k} were used;
then the parameters were allowed to vary and were
determined from our fit. The calculation converged after a few
iterations. 

We varied the input parameters (masses and widths) of the  
$J/\psi$, $\psi(2S)$, and the broad resonances used in the radiative
correction determination
by one standard 
deviation from the values quoted in ref.~\cite{PDG2k}, and find 
that the changes
in the $R$ value are less than 1\% for most points.
Points close to 
the resonance at 4.0 GeV have errors from 1 to 1.7\%. 
Figure~\ref{fig:eff_isr99} (b) shows the radiative correction as a 
function of c.m. energy, where the structure at higher energy is 
related to the 
radiative tail of the $\psi(2S)$ and the broad resonances in this 
energy region.   
Tables~\ref{tab:value} and \ref{tab:error}
list some of the values used in the determination of $R$ and the
contributions to the uncertainty in the value of $R$ at a few
typical energy points in the scanned energy range, respectively.

\begin{table}[!htbp]
\caption{Some values used in the determination of $R$ at a few
typical energy points.}
\begin{tabular}{ccccccccc}
$E_{cm}$ & $N_{had}^{obs}$ & $N_{ll}+$ & $L$  &
$\epsilon(0)$ & $1+\delta_{obs}$ & $R$  & Stat.  & Sys.  \\
(GeV)  & & $N_{\gamma\gamma}$ & (nb$^{-1})$ & (\%) & &  & error & error \\
\tableline
2.000 & 1155.4 & 19.5 &  47.3 & 49.50 & 1.024 & 2.18 & 0.07 & 0.18 \\
3.000 & 2055.4 & 24.3 & 135.9 & 67.55 & 1.038 & 2.21 & 0.05 & 0.11 \\
4.000 &  768.7 & 58.0 &  48.9 & 80.34 & 1.055 & 3.16 & 0.14 & 0.15 \\
4.800 & 1215.3 & 92.6 &  84.4 & 86.79 & 1.113 & 3.66 & 0.14 & 0.19 \\
\end{tabular}
\label{tab:value}
\end{table}

\begin{table}[htbp]
\caption{Contributions to systematic errors: experimental selection of
hadronic events, luminosity determination, theoretical modeling of 
hadronic events,
trigger efficiency, radiative corrections and total systematic error.
All errors are in percentages (\%).}
\begin{tabular}{ccccccc}
$E_{cm}$ & hadron    & $L$ & M.C.     & trigger & radiative  & total \\
 (GeV)   & selection &     & modeling &         & correction &      \\
\tableline
2.000 & 7.07 & 2.81 & 2.62 & 0.5 & 1.06 & 8.13 \\
3.000 & 3.30 & 2.30 & 2.66 & 0.5 & 1.32 & 5.02 \\
4.000 & 2.64 & 2.43 & 2.25 & 0.5 & 1.82 & 4.64 \\
4.800 & 3.58 & 1.74 & 3.05 & 0.5 & 1.02 & 5.14 \\
\end{tabular}
\label{tab:error}
\end{table}

\begin{figure}[!htb] 
\epsfysize=3.5in
\centerline{\epsfbox{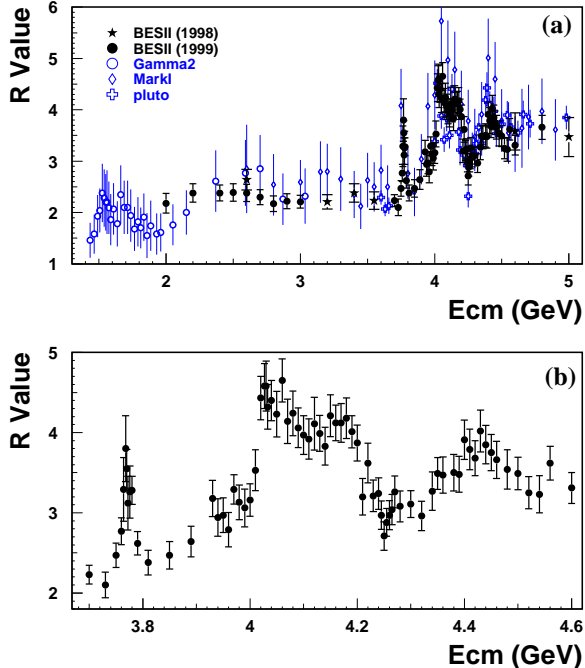}}
\caption{(a) A compilation of measurements of $R$ in the c.m.
energy range from 1.4 to 5 GeV. (b) $R$ values from this experiment 
in the resonance region between 3.7 and 4.6 GeV.} 
\label{fig:besr}
\end{figure}

Table~\ref{tab:rvalue} lists the values of $R$ from this experiment.
They are displayed in 
Fig.~\ref{fig:besr}, together with BESII values from ref.~\cite{besr_1} and 
those measured by MarkI, $\gamma\gamma 2$, and 
Pluto~\cite{mark1,gamma2,pluto}.
The $R$ values from BESII have an average uncertainty of
about 6.6\%, which represents
a factor of two to three improvement in precision 
in the 2 to 5 GeV 
energy region.  Of this error, 3.3\% is common to all points. 
These improved measurements have 
a significant impact on the 
global fit to the electroweak data and
the determination of the SM prediction for the mass
of the Higgs particle \cite{bolekpl}.   In addition, they
are expected to provide an
improvement in the precision of the calculated value of
$a_{\mu}^{SM}$~\cite{ichep2k,martin}, and test the QCD sum rules 
down to 2 GeV~\cite{dave,kuehn}. 

\begin{table*}[htbp]
\begin{center}
\caption{Values of $R$ from this experiment; the first error is 
statistical, the second systematic ($E_{cm}$ in GeV).} 
%
\begin{tabular}{cccccccc}
$E_{cm}$ & $R$   & $E_{cm}$ & $R$ &   $E_{cm}$ & $R$ &
$E_{cm}$ & $R$ \\ \hline
2.000& $2.18\pm0.07\pm0.18$ &3.890& $2.64\pm0.11\pm0.15$ &4.120
& $4.11\pm0.24\pm0.23$ &4.340& $3.27\pm0.15\pm0.18$ \\
2.200& $2.38\pm0.07\pm0.17$ &3.930& $3.18\pm0.14\pm0.17$ &4.130
& $3.99\pm0.15\pm0.17$ &4.350& $3.49\pm0.14\pm0.14$ \\
2.400& $2.38\pm0.07\pm0.14$ &3.940& $2.94\pm0.13\pm0.19$ &4.140
& $3.83\pm0.15\pm0.18$ &4.360& $3.47\pm0.13\pm0.18$ \\
2.500& $2.39\pm0.08\pm0.15$ &3.950& $2.97\pm0.13\pm0.17$ &4.150
& $4.21\pm0.18\pm0.19$ &4.380& $3.50\pm0.15\pm0.17$ \\
2.600& $2.38\pm0.06\pm0.15$ &3.960& $2.79\pm0.12\pm0.17$ &4.160
& $4.12\pm0.15\pm0.16$ &4.390& $3.48\pm0.16\pm0.16$ \\
2.700& $2.30\pm0.07\pm0.13$ &3.970& $3.29\pm0.13\pm0.13$ &4.170
& $4.12\pm0.15\pm0.19$ &4.400& $3.91\pm0.16\pm0.19$ \\
2.800& $2.17\pm0.06\pm0.14$ &3.980& $3.13\pm0.14\pm0.16$ &4.180
& $4.18\pm0.17\pm0.18$ &4.410& $3.79\pm0.15\pm0.20$ \\
2.900& $2.22\pm0.07\pm0.13$ &3.990& $3.06\pm0.15\pm0.18$ &4.190
& $4.01\pm0.14\pm0.14$ &4.420& $3.68\pm0.14\pm0.17$ \\
3.000& $2.21\pm0.05\pm0.11$ &4.000& $3.16\pm0.14\pm0.15$ &4.200
& $3.87\pm0.16\pm0.16$ &4.430& $4.02\pm0.16\pm0.20$ \\
3.700& $2.23\pm0.08\pm0.08$ &4.010& $3.53\pm0.16\pm0.20$ &4.210
& $3.20\pm0.16\pm0.17$ &4.440& $3.85\pm0.17\pm0.17$ \\
3.730& $2.10\pm0.08\pm0.14$ &4.020& $4.43\pm0.16\pm0.21$ &4.220
& $3.62\pm0.15\pm0.20$ &4.450& $3.75\pm0.15\pm0.17$ \\
3.750& $2.47\pm0.09\pm0.12$ &4.027& $4.58\pm0.18\pm0.21$ &4.230
& $3.21\pm0.13\pm0.15$ &4.460& $3.66\pm0.17\pm0.16$ \\
3.760& $2.77\pm0.11\pm0.13$ &4.030& $4.58\pm0.20\pm0.23$ &4.240
& $3.24\pm0.12\pm0.15$ &4.480& $3.54\pm0.17\pm0.18$ \\
3.764& $3.29\pm0.27\pm0.29$ &4.033& $4.32\pm0.17\pm0.22$ &4.245
& $2.97\pm0.11\pm0.14$ &4.500& $3.49\pm0.14\pm0.15$ \\
3.768& $3.80\pm0.33\pm0.25$ &4.040& $4.40\pm0.17\pm0.19$ &4.250
& $2.71\pm0.12\pm0.13$ &4.520& $3.25\pm0.13\pm0.15$ \\
3.770& $3.55\pm0.14\pm0.19$ &4.050& $4.23\pm0.17\pm0.22$ &4.255
& $2.88\pm0.11\pm0.14$ &4.540& $3.23\pm0.14\pm0.18$ \\
3.772& $3.12\pm0.24\pm0.23$ &4.060& $4.65\pm0.19\pm0.19$ &4.260
& $2.97\pm0.11\pm0.14$ &4.560& $3.62\pm0.13\pm0.16$ \\
3.776& $3.26\pm0.26\pm0.19$ &4.070& $4.14\pm0.20\pm0.19$ &4.265
& $3.04\pm0.13\pm0.14$ &4.600& $3.31\pm0.11\pm0.16$ \\
3.780& $3.28\pm0.12\pm0.12$ &4.080& $4.24\pm0.21\pm0.18$ &4.270
& $3.26\pm0.12\pm0.16$ &4.800& $3.66\pm0.14\pm0.19$ \\
3.790& $2.62\pm0.11\pm0.10$ &4.090& $4.06\pm0.17\pm0.18$ &4.280
& $3.08\pm0.12\pm0.15$ &     &                      \\
3.810& $2.38\pm0.10\pm0.12$ &4.100& $3.97\pm0.16\pm0.18$ &4.300
& $3.11\pm0.12\pm0.12$ &     &                      \\
3.850& $2.47\pm0.11\pm0.13$ &4.110& $3.92\pm0.16\pm0.19$ &4.320
& $2.96\pm0.12\pm0.14$ &     &                      \\
\end{tabular}
\label{tab:rvalue}
\end{center}
\end{table*}



We would like to thank the staff of the BEPC Accelerator Center
and IHEP Computing Center for their efforts.  
We thank B. Andersson for helping in the development of the LUARLW generator.
We also wish to acknowledge useful discussions with M. Davier, 
B. Pietrzyk, T. Sj\"{o}strand,  A. D. Martin and M. L. Swartz.
We especially thank M. Tigner for major contributions not only to
BES but also to the operation of the BEPC during the $R$ scan.

This work is supported in part by the National Natural
 Science Foundation of China under Contract Nos. 19991480,
19805009 and 19825116; the Chinese
 Academy of Sciences under contract Nos. KJ95T-03, and E-01 (IHEP);
 and by the Department of
 Energy under Contract Nos.
 DE-FG03-93ER40788 (Colorado State University),
 DE-AC03-76SF00515 (SLAC),
 DE-FG03-94ER40833 (U Hawaii), DE-FG03-95ER40925 (UT Dallas),
and by the Ministry of Science and Technology of Korea under Contract
KISTEP I-03-037(Korea).



\begin{thebibliography}{99}
\bibitem{rlowe} Z.G. Zhao, International Journal of Modern Physics A15
(2000)3739.
\bibitem{mark1} J. L. Siegrist {\it et al.}, (Mark I Collab.), \Journal
{\PRD}{26}{969}{1982}.         
\bibitem{gamma2} C. Bacci {\it et al.}, ($\gamma \gamma2$ Collab.),
\Journal{\PLB}{86}{234}{1979}.
\bibitem{pluto} L. Criegee and G. Knies, (Pluto Collab.),
{\em Phys. Rep.} {\bf 83}, 151 (1982);\\
Ch. Berger {\it et al.}, \Journal {\PLB}{81}{410}{1979}.
\bibitem{rscan2} Z.G. Zhao, \Journal{\NPA}{675}{13c}{2000}. 
\bibitem{bes2}J.Z. Bai {\it et al.}, (BES Collab.),
\Journal{\NIM}{A458}{627}{2001}.
\bibitem{besr_1} J. Z. Bai {\it et al.}, (BES Collab.),
\Journal{\PRL}{84}{594}{2000}.
\bibitem{bo} B. Andersson and Haiming Hu, ``Few-body States in Lund String
Fragmentation Model'', hep-ph/9910285.
\bibitem{eichiten}E. Eichten {\it et al.}, \Journal{\PRD}{21}{203}{1980}.
\bibitem{chenjc}J.C. Chen {\it et al.}, \Journal{\PRD}{62}{034003}{2000}.
\bibitem{huhm} Haiming Hu {\it et al.}, High Energy Physics and Nuclear
Physics (in Chinese), 25, 1035(2001)
\bibitem{berends}F.A. Berends and R. Kleiss, \Journal{\NPB}{178}{141}{1981}.
\bibitem{fmartin}G. Bonneau and F. Martin, \Journal{\NPB}{27}{387}{1971}.
\bibitem{fadin}E. A. Kuraev and V.S. Fadin, {\em Sov. J. Nucl.
Phys.} {\bf41}, 3(1985). 
\bibitem{crystalball}A. Osterheld {\it et al.}, SLAC-PUB-4160, 1986. (T/E)
\bibitem{comment}
Haiming Hu {\it et al.}, High Energy Physics and Nuclear Physics
(in Chinese), 25, 701(2001)
\bibitem{PDG2k} Particle Data Group, D.E. Groom {\it et al.}, Eur. Phys. J.
C15, 1 (2000).
\bibitem{bolekpl} H. Burkhardt and B. Pietrzyk,
\Journal{\PLB}{513}{46}{2001}.
\bibitem{ichep2k} B. Pietrzyk, Robert Carey, Atul Gurtu, 
talks given at ICHEP2000, Osaka, Japan, July 2000.
\bibitem{martin} A. Martin {\it et al.}, \Journal{\PLB}{492}{69}{2000}. 
\bibitem{dave} M. Davier and A. Hoecker, \Journal{\PLB}{419}{419}{1998}.
\bibitem{kuehn} J.H. Kuehn and M. Steinhauser, \Journal{\PLB}{437}{425}{1998}.

\end{thebibliography}
\end{document}